\documentclass[10pt,twocolumn]{article}
\usepackage[textwidth=7in,textheight=8.5in]{geometry}
\usepackage{graphicx}
\usepackage{amsmath} 
\usepackage{amssymb} 
\usepackage{fancyvrb} 
\usepackage{soul}
\usepackage{fancyhdr}
\usepackage{epstopdf}
\usepackage{lineno}
\makeatletter
\renewcommand{\maketitle}{\bgroup
\begin{flushleft}
  \begin{Huge}
  \textbf{\@title}\\
  \end{Huge}
  \vspace{1cm}
  \@author
\end{flushleft}\egroup
}
\makeatother
\title{Fluctuation of USA Gold Price - Revisited with Chaos-based Complex Network Method}
\author{%
    \textbf{{\large Susmita Bhaduri}}$^{1}$, \textbf{{\Large Dipak Ghosh}}$^{2}$, \textbf{{\Large Subhadeep Ghosh}}$^{3}$\\
    $^{1}$Deepa Ghosh Research Foundation, Kolkata-700031,India \\
    $^{2}$Deepa Ghosh Research Foundation, Kolkata-700031,India \\
    $^{3}$Deepa Ghosh Research Foundation, Kolkata-700031,India \\
    $^{3}$432E 88 Street 505 New York 128 USA \\
    \underline{$^{1}$susmita.sbhaduri@dgfoundation.in}\\
    \underline{$^{2}$deegee111@gmail.com}\\
    \underline{$^{3}$subhadeepghosh81@gmail.com}
}
\begin{document}
\twocolumn[
  \begin{@twocolumnfalse}
    \maketitle
  \end{@twocolumnfalse}
  ]
\noindent


%

%
%

\date{\today}

\begin{abstract}
We give emphasis on the use of chaos-based rigorous non-linear technique called Visibility Graph Analysis, to study one economic time series - gold price of USA. This method can offer reliable results with \textbf{finite} data. This paper reports the result of such an analysis on the times series depicting the fluctuation of gold price of USA for the span of 25 years($1990 - 2013$). 
This analysis reveals that a quantitative parameter from the theory can explain satisfactorily the real life nature of fluctuation of gold price of USA and hence building a strong database in terms of a quantitative parameter which can eventually be used for forecasting purpose. 
\end{abstract}




\section{Introduction}
\label{intro}
In the modern science of finance, the application of financial physics has been developed recently.
Many studies have found the financial time series to exhibit some non-linear properties such as long-memory in volatility~\cite{liu1990,yam2005,oh2006,wang2006}, a Multi-fractal nature~\cite{pari1985,iva1999,muz2000,cal2001,eis2004,sai2006,bar2010}, and fat tails~\cite{mant1995,mant1996,ple2003,gab2003}. 
Extensive methods have been adopted to extract the empirical Multi-fractal properties in financial data sets, for instance, the Wavelet Transform Module Maxima [WTMM] as per~\cite{hols1988,muz1991,muz2000} and the Multi-fractal Detrended Fluctuation Analysis (MF-DFA)~\cite{kant2002}.

Self-similar processes such as fractional Brownian motion (fBm) are currently used to model fractal phenomena of different nature, ranging from physics, biology, economics or engineering. fBm has been used in models of electronic de-localization, as a theoretical framework to analyze turbulence data, to describe geologic properties, to quantify correlations in DNA base sequences, to characterize physiological signals such as ECG, EEG, network traffic and even to categorize music signal emotion-wise. Fractional Brownian motion $B_{H}(t)$ is a non-stationary random process with stationary self-similar increments (fractional Gaussian noise, fGn) that can be characterized by the Hurst exponent($H$), where $0 < H < 1$. The one-step memory Brownian motion is obtained for $H = 1/2$ , whereas time series with $H > 1/2$  shows persistence and anti-persistence if $H < 1/2$.

Though Hurst exponent has been used extensively for financial analysis and it has successfully detected long range correlations in financial time-series, its computation is still a problem. The main problem about the Hurst exponent is the effect of \textbf{finite} data length for the estimation of Hurst exponent. The Hurst exponent yields to most precise and accurate result for random processes such as Brownian motion time-series with an \textbf{infinite} number of data points. But in real life situation we use finite time-series to estimate the Hurst exponent, long-range correlations in the time-series are partially broken into finite series and local dynamics corresponding to a particular temporal window are overestimated. Hence, the Hurst exponent calculated for real financial data inevitably deviates from its real value.

It is a common practise to use MF-DFA technique for this type of analysis due to its obvious advantage of having highest precision in the scaling analysis.
However, as has been discussed earlier this method suffers from one lacuna. This theory demands that the length of the time-series to be analysed has to be infinite, whereas in real life this time-series is always \textbf{‘finite’} because there is no other option. 
In this regard another radically different rigorous methods- \textbf{Visibility network analysis} - has been reported  by Lacasa et.al.~\cite{laca2008,laca2009}. Recently this method is extensively used over \textbf{finite} time-series data set and has produced reliable result in several domains of science and social science.

Lacasa et.al.~\cite{laca2008,laca2009} has introduced the visibility algorithm, based on graph theoretical techniques. A visibility graph is obtained from the mapping of a time series into a network. As already mentioned the advantage of visibility graph technique is that it gives more accurate estimate of Hurst exponent compared to other method(MFDFA) since MFDFA theory demands that the calculation must be done on infinite series, but in practise it is done on fine time series, resulting in wrong estimation of Hurst exponent. This method is very much suitable for analysis \textbf{finite} time series (real life situation). The reliability of this novel and new methodology is confirmed with exhaustive numerical simulations as well as with analytical developments~\cite{laca2008,laca2009}. 

It has been found from empirical study that during financial deregulation the stock markets of a country become sensitive to both domestic and peripheral financial factors. One such factor is gold price. Gold has been used as money and as a relative standard for currency throughout history. 
The price of gold and stock fluctuates in an opposite direction, globally. People decrease their investment in gold when its price is low and increase the same for the case of stock price. This process increases the value of stock price due to this huge investment.
Also, when the stock prices are low, people invest more in gold while waiting for the crisis to fade away. This again increase the demand for gold and in turn the price of gold. In effect gold is a substitute investment option for investors.
When the gold price is in rising trend, investors go for investing in gold and lessen investment in stock market. This makes stock prices to fall. Hence we can expect \textbf{an negative relationship between gold and stock price}~\cite{sarba2013}. Gaur et al.~\cite{gaur2010} also documented the historical evidences on the simultaneous fluctuation of gold price and stock price in India. He also concluded that when the stock market crashes or when the dollar
weakens, gold continues to be a safe haven investment because gold prices rise in such circumstances.

Recently a few works have been reported where interesting attempts to study fluctuation of USA gold price as well as Indian stock market BSE using MF-DFA technique. However, as briefed above due to small sample size the results may not be reliable as expected from the theory. Also Yu Long et al.~\cite{yu2013} have mapped the gold price time series into a visibility graph network, and have explored the mechanism underlying the gold price fluctuation from the perspective of complex network theory and also analysed the nature of the gold price fluctuation. In view of this in the present investigation we propose to perform analysis using \textbf{Visibility graph technique} with prime objective of 
\begin{enumerate}
\item Studying the fluctuation pattern of USA gold price.
\item Analysing scope of the application of quantitative visibility graph analysis as a pre-cursor of financial crisis of course with proper validation.
\end{enumerate}

The rest of the paper is organized as follows. The method of analysis is explained in the Section~\ref{ana}, then in Section~\ref{data} the details of data is elaborated. The result is analysed and the inferences from the test results are presented in Section~\ref{res}. Finally the paper is concluded in Section~\ref{con}.

\section{Method of analysis}
\label{ana}
We would briefly describe the \textbf{Visibility graph technique} in this section. 

\subsection{Visibility Graph Algorithm}
\begin{figure}
\includegraphics[scale=.4]{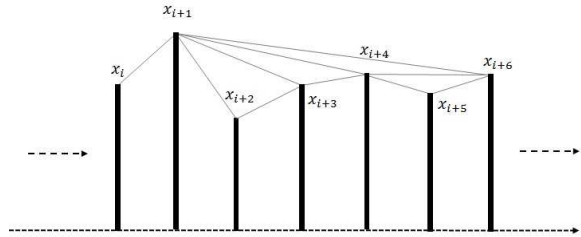}
\caption{Visibility Graph for time series X}
\label{visi}
\end{figure}
The visibility graph algorithm maps time series $X$ to its Visibility Graph. Suppose the $i^th$ point of the time series, $X_{i}$. Two vertices (nodes) of the graph,$X_{m}$ and $X_{n}$, are connected via a bidirectional edge if and only if the below equation is valid.
\setcounter{equation}{0}
\begin{equation}
X_{m+j} < X_{n} + (\frac{n - (m+j)}{n-m})\cdot(X_m - X_n) 
\label{ve}
\end{equation}
\begin{math}
\mbox{where }
\forall j \in Z^{+} \mbox{ and } j < (n-m)\\
\\
\end{math} 
In Fig.\ref{visi} it is shown that $X_{m}$ and  $X_{n}$ can see each other if the Eq. \ref{ve} is valid.
As per the VG algorithm two sequential points of the time series can see each other hence all sequential nodes are connected together. 

\textbf{Note:We should be converting the time series to positive planes as the above algorithm is valid for positive $X$ values in the time series.}

\subsubsection{Power of Scale-freeness of VG \mbox{-} PSVG}
The degree of a node in a graph \mbox{-} here VG is the number of connections or edges the node has to other nodes. The degree distribution $P(k)$ of a network is then defined to be the fraction of nodes with degree $k$ in the network. Thus if there are $n$ nodes in total in a network and $n_k$ of them have degree $k$, we have $P(k) = n_k/n$.

A power law is a functional relationship between two quantities, where one quantity varies as a power of another. The scale-freeness property of Visibility Graph states that the degree distribution of it's nodes follow \textbf{Power Law} :$P(k) \sim k^{-\lambda_p}$, where $\lambda_p$ is a constant and it is called the \textbf{power of the scale-freeness}. 

As per Lacasa et al.~\cite{laca2008,laca2009} the power of the scale-free structure $(\lambda_p)$  of the VG corresponds to the amount of fractality of the time series,
and the slope of $log_2[P(k)]$ versus $log_2[1/k]$, indicates the FD \mbox{-} Fractal Dimension of the signal. This value of the slope known as Power of Scale-freeness in Visibility Graph (PSVG) as a measure of complexity and fractality of the time series. PSVG is denoted by $\lambda_p$ here.

\subsection{Our analysis}

\begin{figure}
\includegraphics[width=3in]{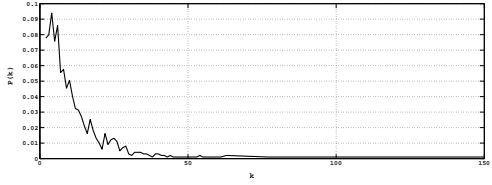}
\caption{$P(k)$ vs $k$ for USA gold price fluctuation for timespan 1998-2001 }
\label{gold}
\end{figure}

First we have constructed visibility graph from the time series datasets of USA Gold price fluctuation and verified whether they are confirming the power law. The graph of $P(k)$ is plotted against $k$ for the $6$ sets of USA gold price fluctuation data. An example graph is shown in Fig~\ref{gold}.

\begin{figure}
\includegraphics[width=3in]{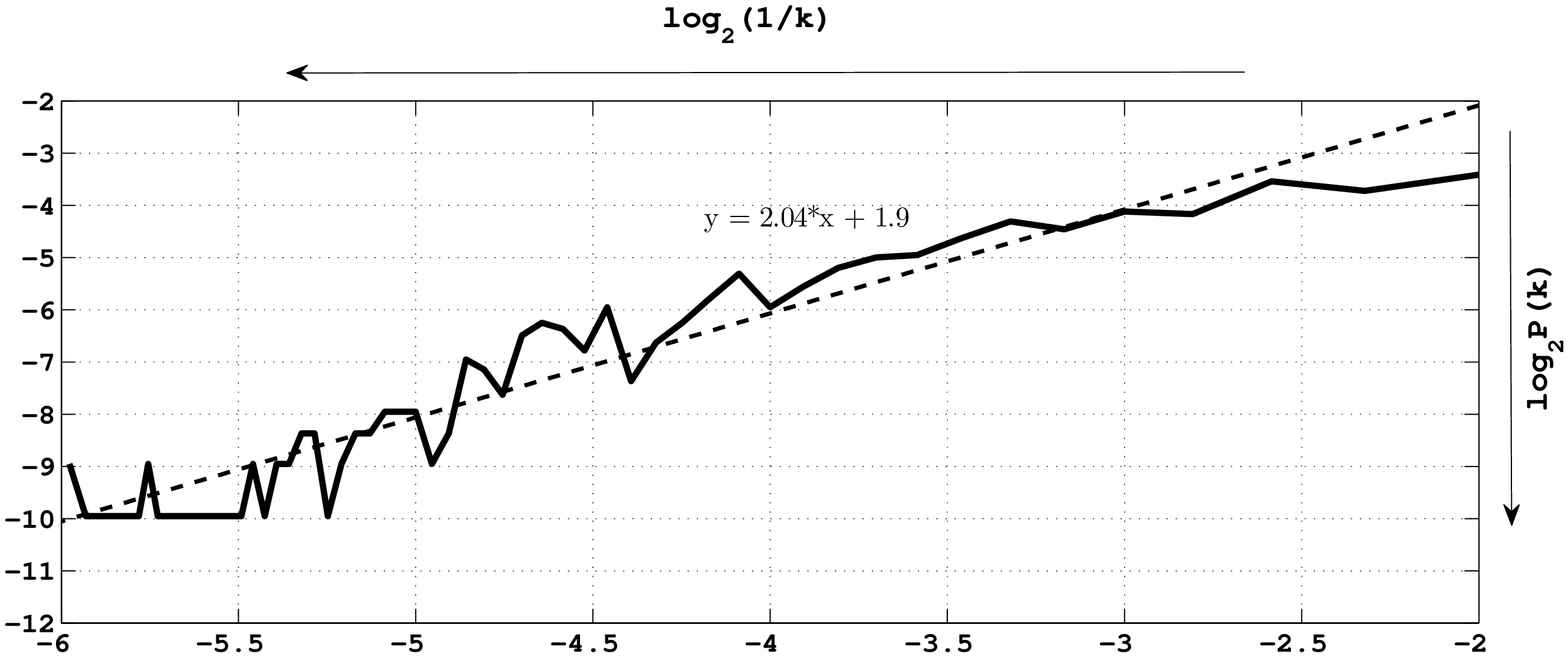}
\caption{$log_2P(k)$ vs $log_2(1/k)$ for USA Gold price fluctuation for timespan 1998-2001 }
\label{goldf}
\end{figure}

As trend of $P(k)$ w.r.t. $k$ is confirming to the \textbf{Power Law}, Power of Scale-freeness in Visibility Graph (PSVG) is calculated for the graph from the slope of $log_{2}[P(k)]$ versus $log_{2}[1/k]$ as shown in the Fig.~\ref{goldf}. 

The slope is denoted by $\lambda_{pg}$ for USA Gold price fluctuation. 
Then we have analysed the trend of $\lambda_{pg}$ for all the dataset and inferences drawn from there.

\section{Data description}
\label{data}

We have taken the datasets for USA gold price fluctuation from www.usagold.com for the span of $1990-2013$. We have divided the time series into subsets: (i) Jan 1990-Dec 1993, (ii) Jan 1994-Dec 1997, (iii) Jan 1998-Dec
2001, (iv) Jan 2002-Dec 2005, (v) Jan 2006-Dec 2009, (vi) Jan 2010-May 2013.

\begin{figure}
\includegraphics[width=3in]{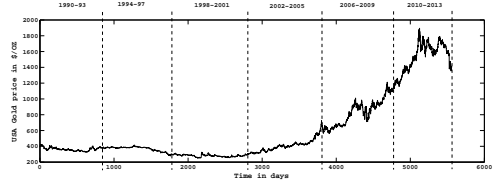}
\caption{USA gold price in \$/OZ in the period Jan 1990-May 2013.}
\label{goldata}
\end{figure}

Fig.\ref{goldata} shows the fluctuation of USA gold price, for the period Jan 1990-May 2013.

\section{Results}
\label{res}
\begin{figure*}
\includegraphics[width=6in]{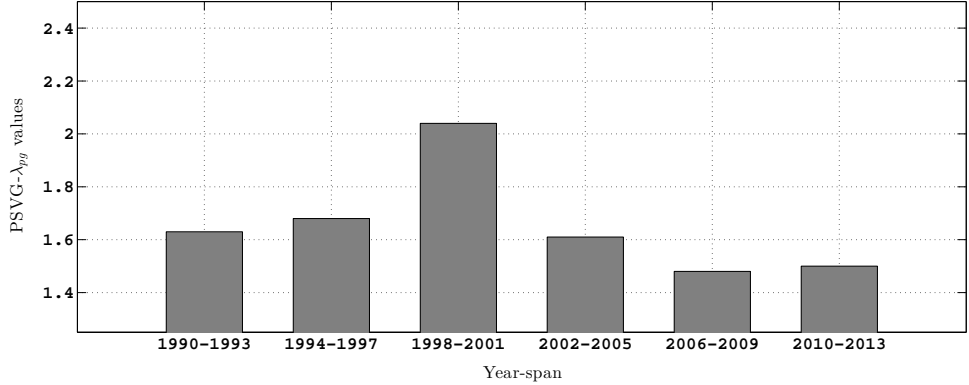}
\caption{Comparison of $\lambda_{pg}$ values for USA gold price fluctuation}
\label{trend}
\end{figure*}

The Fig.~\ref{trend} shows the year-span wise comparison of PSVG-s for USA gold price fluctuation($PSVG = \lambda_{pg}$).
The Table~\ref{result} shows the values of $\lambda_{pg}$ calculated for the fluctuation of USA gold price.

\begin{table}
\caption{Year span-wise comparison of $\lambda_{pg}$ calculated for the fluctuation of USA gold price}
\label{result}
\begin{tabular}{|c|c|} 
\hline
\textbf{Year-span}&\textbf{$\lambda_{pg}$}\\
\hline
1990-1993&$1.63$\\
\hline
1994-1997&$1.68$\\
\hline
1998-2001&$2.04$\\
\hline
2002-2005&$1.61$\\
\hline
2006-2009&$1.48$\\
\hline
2010-2013&$1.50$\\
\hline
\end{tabular} 
\end{table}

The results of our analysis of the values of $\lambda_{pg}$ from the Table.~\ref{result} and Fig.~\ref{trend}, are described below.
\begin{enumerate}
\item It is interesting to observe that PSVG for USA gold price fluctuation($PSVG = \lambda_{pg}$) assumes maximum value during (1998-2001) and minimum value during the year span (2006-2009) and also the value is very low and nearly minimum, during (2010-2013). 

\item If one examines the real scenario of gold price of USA during the period under investigation, it is observed that the gold price(in \$/OZ) was minimum during the period (1998-2001) and the fluctuation of the price is also insignificant. The obvious interpretation is a \textbf{high} value of PSVG 
assumes \textbf{low} rate of fluctuation of gold price along with the exact price also.

The same behaviour is also reflected if one conforms the PSVG value with gold price during the period (2006-2009) and also (2010-2013), when PSVG values assume \textbf{low} values including the minimum one. The gold price(in \$/OZ) during those period is not only significantly \textbf{high}, the fluctuation rate is also on the higher side.

\item Thus this analysis clearly manifests one important and significant information about and \textbf{inverse relationship} between PSVG values and gold price(in \$/OZ) in USA, including its fluctuation.

\item This remarkable agreement of PSVG values and real life scenario prompts us to propose that the PSVG parameter may be used as a reliable pre-cursor for instability of the financial market. We are further encouraged to suggest to analyse further data and the trend of PSVG parameter of gold price, can be studied in detail continuously to forecast financial crisis since it has been globally accepted that the price of gold and stock has a negative relationship~\cite{sarba2013}.
Validation can then easily be done with the help of real life scenario as and when it becomes available.

\end{enumerate}

The observation encourages us for further refinement of the analysis including more and more data from other country and further subdivision of duration of analysis.

\section{Conclusion} 
\label{con}
Since the visibility graph technique gives the most reliable result even with very \textbf{short-length and finite} time-series we have re-visited the study of USA gold price fluctuation to extract most reliable results in terms of quantitative parameters for the first time, so far as our knowledge goes.
To conclude we may highlight the importance of present investigation precisely in the following:
\begin{enumerate}
\item Our analysis manifests satisfactory agreement between the PSVG parameter and the real life scenario of USA gold price fluctuation.

\item We may analyze with detailed data of gold prices of different countries. The continuous of the year wise trend of the PSVG for gold price fluctuation($PSVG = \lambda_{pg}$) may be used as a database for forecasting economic crisis.

\end{enumerate}

This study encourages further similar analysis of different financial series using this technique which may yield to a optimal methodology for tackling risk management in investment.
Finally we emphasize that the chaos based random fractal techniques may prove itself more appropriate method of analyzing financial time series with a possibility of forecasting.

\section{References}
\bibliographystyle{splncs} 
\bibliography{goldarXiv}

\begin{thebibliography}{10}

\bibitem{liu1990}
Y.Liu.
\newblock Physical Review E \textbf{60} (1990)  1390

\bibitem{yam2005}
K.Yamasaki, ed.
\newblock Volume 102 of Proceedings of the National Academy of Sciences of the
  United States of America. (2005)

\bibitem{oh2006}
G.Oh.
\newblock Journal of the Korean physical Society \textbf{48} (2006)  197

\bibitem{wang2006}
F.Wang.
\newblock Physical Review E \textbf{72} (2006)  066128

\bibitem{pari1985}
G.Parisi, U.Grisch, eds.:
\newblock Turbulence and Predictability in Geophysical Fluid Dynamics and
  Climate Dynamics.
\newblock Proceedings of the International School (Enrico Fermi)
  (Northy-Holland, Amsterdam, Netherlands, 1985). (1985)

\bibitem{iva1999}
P.Ch.Ivanov.
\newblock Nature \textbf{399} (1999)  461

\bibitem{muz2000}
J.F.Muzy.
\newblock European Physical Journal B \textbf{17} (2000)  537

\bibitem{cal2001}
L.Calvet.
\newblock Journal of Econometrics \textbf{105} (2001) ~27

\bibitem{eis2004}
Z.Eisler.
\newblock Physica A \textbf{343} (2004)  603

\bibitem{sai2006}
A.Saichev, D.Sornette.
\newblock Physical Review E \textbf{74} (2006)  011111

\bibitem{bar2010}
J.Barunik, L.Kristoufek.
\newblock Physica A \textbf{389} (2010)  3844

\bibitem{mant1995}
R.N.Mantegna.
\newblock Nature \textbf{376} (1995) ~46

\bibitem{mant1996}
R.N.Mantegna.
\newblock Nature \textbf{383} (1996)  587

\bibitem{ple2003}
V.Plerou.
\newblock Nature \textbf{421} (2003)  130

\bibitem{gab2003}
X.Gabaix.
\newblock Nature \textbf{423} (2003)  267

\bibitem{hols1988}
M.Holschneider.
\newblock Journal of Statistical Physics \textbf{50} (1988)  953

\bibitem{muz1991}
J.F.Muzy.
\newblock Physical Review Letters \textbf{67} (1991)  3515

\bibitem{kant2002}
J.W.Kantelhardt.
\newblock Physica A \textbf{316} (2002) ~87

\bibitem{laca2008}
L.Lacasa.
\newblock Proc. Matl. Acad. Sci. U.S.A. \textbf{105} (2008)

\bibitem{laca2009}
L.Lacasa.
\newblock Europhys. Lett. \textbf{86} (2009)

\bibitem{sarba2013}
Ray, S.
\newblock Causal nexus between gold price movement and stock market: evidence
  from indian stock market, Sciknow Publications Ltd. Econometrics, Attribution
  3.0 Unported (CC BY 3.0) \textbf{1(1)} (2013)  12--19

\bibitem{gaur2010}
Gaur, A., Bansal, M.
\newblock A comparative study of gold price movements in Indian and global
  markets, Indian J. Finance \textbf{4(2)} (2010)  32--37

\bibitem{yu2013}
Long, Y.
\newblock Visibility graph network analysis of gold price time series,Physica
  A: Statistical Mechanics and its Applications \textbf{392(16)} (2013)
  3374--3384

\end{thebibliography}
 
\end{document}